\documentclass[useAMS,usenatbib,usegraphicx]{mn2e}

\title[Three Planets in HD 11964]{A Bayesian Periodogram Finds Evidence for Three Planets in HD 11964}
\author[P. C. Gregory]{P. C. Gregory$^{1}$\thanks{E-mail:
gregory@phas.ubc.ca}\footnotemark[1]\thanks{http://www.physics.ubc.ca/~gregory/gregory.html}\\
$^{1}$Physics and Astronomy Department, University of British Columbia,\\ 6224 Agricultural Rd., Vancouver, British Columbia, V6T 1Z1, Canada}
\begin{document}

\date{MNRAS in press Aug. 7, 2007}

\pagerange{\pageref{firstpage}--\pageref{lastpage}} \pubyear{2006}

\maketitle

\label{firstpage}

\begin{abstract}
A Bayesian multi-planet Kepler periodogram has been developed for the analysis of precision radial velocity data (Gregory 2005b and 2007). The periodogram employs a parallel tempering Markov chain Monte Carlo algorithm. The HD 11964 data \citep{Butler2006} has been re-analyzed using 1, 2, 3 and 4 planet models. Assuming that all the models are equally probable {\it a priori}, the three planet model is found to be $\ge 600$ times more probable than the next most probable model which is a two planet model. The most probable model exhibits three periods of $38.02_{-0.05}^{+0.06}$, $360_{-4}^{+4}$ and $1924_{-43}^{+44}$ d, and eccentricities of $0.22_{-0.22}^{+0.11}$, $0.63_{-0.17}^{+0.34}$ and $0.05_{-0.05}^{+0.03}$, respectively. Assuming the three signals (each one consistent with a Keplerian orbit) are caused by planets, the corresponding limits on planetary mass ($M \sin i$) and semi-major axis are \\ 
$(0.090_{-0.14}^{+0.15} M_J, 0.253_{-0.009}^{+0.009}\rm{au}), (0.21_{-0.07}^{+0.06} M_J, 1.13_{-0.04}^{+0.04}\rm{au}), (0.77_{-0.08}^{+0.08} M_J, 3.46_{-0.13}^{+0.13}\rm{au})$,\\ respectively. The small difference ($1.3 \sigma$) between the 360 day period and one year suggests that it might be worth investigating the barycentric correction for the HD 11964 data. 
\end{abstract}


\section{Introduction}
\label{sec:introduction}

Improvements in precision radial velocity measurements and continued monitoring are permitting the detection of lower amplitude planetary signatures. One example of the fruits of this work is the detection of a super earth in the habital zone surrounding Gliese 581 \citep{Urdy2007}. This and other remarkable successes on the part of the observers is motivating a significant effort to improve the statistical tools for analyzing radial velocity data (e.g., \citealt{FordGregory2006}, Ford 2005 \& 2006, \citealt{Gregory2005b}, \citealt{Cumming2004}, \citealt{LoredoChernoff2003}, \citealt{Loredo2004}). Much of the recent work has highlighted a Bayesian MCMC approach as a way to better understand parameter uncertainties and degeneracies and to compute model probabilities.

Gregory (2005a, b \& c and 2007) presented a Bayesian MCMC algorithm that makes use of parallel tempering to efficiently explore a large model parameter space starting from a random location. It is able to identify any significant periodic signal component in the data that satisfies Kepler's laws and thus functions as a Kepler periodogram~\footnote{Following on from the pioneering work on Bayesian periodograms by \citet{Jaynes1987} and \citet{Brett1988}}. This eliminates the need for a separate periodogram search for trial orbital periods which typically assume a sinusoidal model for the signal that is only correct for a circular orbit. In addition, the Bayesian MCMC algorithm provides full marginal parameters distributions for all the orbital elements that can be determined from radial velocity data. The samples from the parallel chains can also be used to compute the marginal likelihood for a given model \citep{Gregorybook} for use in computing the Bayes factor that is needed to compare models with different numbers of planets. The parallel tempering MCMC algorithm employed in this work includes an innovative two stage adaptive control system that automates the selection of efficient Gaussian parameter proposal distributions. The annealing of the proposal distributions carried out by the control system combined with parallel tempering makes it practical to attempt a blind search for multiple planets simultaneously. This was done for the analysis of the current data set and for the analysis of the HD 208487 reported earlier \citep{Gregory2007}.

This paper presents a Bayesian re-analysis of the existing 87 precision radial velocity measurements for HD 11964 published by \citet{Butler2006}, who reported the detection of a single planet with a period of $2110\pm270$d after removing a trend in the data. They remark that the 5.3m s$^{-1}$ residuals are comparable to the 9m s$^{-1}$ amplitude, placing the exoplanetary interpretation of the velocity variations somewhat in doubt. 

\section{Analysis}

The analysis of the HD 11964 data employed exactly the same Bayesian multi-planet Kepler periodogram that was previously described for the analysis of HD 208487 \citep{Gregory2007}. The periodogram utilizes a parallel tempering Markov chain Monte Carlo algorithm which yields the probability density distribution for each model parameter and permits a direct comparison of the probabilities of models with differing numbers of planets. In parallel tempering each chain corresponds to a different temperature. We parameterize the temperature by its reciprocal, $\beta=1/{\cal T}$ which varies from zero to 1. For parameter estimation purposes 12 chains ($\beta=\{0.05, 0.1, 0.15, 0.25, 0.35, 0.45, 0.55, 0.65, 0.70, 0.80, 0.90, 1.0\}$) were employed and the final samples drawn from the $\beta  = 1$ chain, which corresponds to the desired target probability distribution. For $\beta \ll 1$, the distribution is much flatter. 

At intervals, a pair of adjacent chains on the tempering ladder are chosen at random and a proposal made to swap their parameter states. The mean number of iterations between swap proposals was set $= 8$. A Monte Carlo acceptance rule determines the probability for the proposed swap to occur. This swap allows for an exchange of information across the population of parallel simulations. In the higher temperature simulations, radically different configurations can arise, whereas in higher $\beta$ (lower temperature) states, a configuration is given the chance to refine itself.

The samples from hotter simulations were also used to evaluate the marginal (global) likelihood needed for model selection, following Section 12.7 of \citet{Gregorybook} and \citet{Gregory2007}. This is discussed more in Section~\ref{sec:modsel}. Marginal likelihoods estimated in this way require many more parallel simulations. For HD 11964, 40 $\beta$ levels were used spanning the range $\beta = 10^{-8}$ to $1.0$ with a mean interval between swaps = 3. 

For a one planet model the predicted radial velocity is given by
\begin{equation}
v(t_i) = V + K [\cos\{\theta(t_i+\chi P)+\omega\} + e \cos \omega],
\label{eq:orbit1}
\end{equation}
and involves the 6 unknown parameters
\begin{itemize}
\item[] $V =$ a constant velocity.
\item[] $K =$ velocity semi-amplitude. 
\item[] $P =$ the orbital period.
\item[] $e =$ the orbital eccentricity.
\item[] $\omega =$ the longitude of periastron.
\item[] $\chi =$ the fraction of an orbit, prior to the start of data taking, that periastron occurred at. Thus, $\chi P =$ the number of days prior to $t_i = 0$ that the star was at periastron, for an orbital period of P days. 
\item[] $\theta(t_i+\chi P) =$ the angle of the star in its orbit relative to periastron at time $t_i$, also called the true anomaly.
\end{itemize}

We utilize this form of the equation because we obtain the dependence of $\theta$ on $t_i$ by solving the conservation of angular momentum equation
\begin{equation}
\frac{d\theta}{dt} - \frac{2 \pi [1+e\cos \theta(t_i+\chi \; P)]^2}{P (1-e^2)^{3/2}} = 0.
\label{eq:orbit2}
\end{equation}
Our algorithm is implemented in {\it Mathematica} and it proves faster for {\it Mathematica} to solve this differential equation than solve the equations relating the true anomaly to the mean anomaly via the eccentric anomaly. {\it Mathematica} generates an accurate interpolating function between $t$ and $\theta$ so the differential equation does not need to be solved separately for each $t_i$. Evaluating the interpolating function for each $t_i$ is very fast compared to solving the differential equation, so the algorithm should be able to handle much larger samples of radial velocity data than those currently available without a significant increase in computational time.

As described in more detail in \citealt{Gregory2007}, we employed a re-parameterization of $\chi$ and $\omega$ to improve the MCMC convergence speed motivated by the work of Ford (2006). The two new parameters are $\psi=2\pi\chi+\omega$ and $\phi=2 \pi\chi-\omega$. $\psi$ is well determined for all eccentricities. Although $\phi$ is not well determined for low eccentricities, it is at least orthogonal to the $\psi$ parameter. In \citealt{Gregory2007}, we recommended a uniform prior for $\psi$ in the interval 0 to $2 \pi$ and uniform prior for $\phi$ in the interval $-2 \pi$ to $+2 \pi$, which is the smallest rectangle in $(\psi,\phi)$ that uniformly samples the full range in $(\chi,\omega)$. However, a posterior that is a wraparound continuous in $(\chi,\omega)$ does not map into a wraparound continuous distribution in this rectangle of $(\psi,\phi)$. This can reduce the algorithms performance and convergence. A simple fix is to double the range of $\psi$ to $(0 < \psi < 4\pi)$.  The big $(\psi,\phi)$ square holds two copies of the probability patch in $(\chi,\omega)$ which doesn't matter. What matters is that the posterior is now wraparound continuous in $(\psi,\phi)$. 

In a Bayesian analysis we need to specify a suitable prior for each parameter. These are tabulated in Table~\ref{tab:priors}. Detailed arguments for the choice of each prior were given in \citealt{Gregory2007}. The lower bound on the search period of 1.01d was employed to avoid a possible 1 day sampling artifact.    
\begin{table*}
 \centering
 \begin{minipage}{140mm}
  \caption{Prior parameter probability distributions.}
  \label{tab:priors}
  \begin{tabular}{@{}llll@{}}
  \hline
   Parameter    &    Prior        & Lower bound & Upper bound\\
 \hline
Orbital frequency  & $p(\ln f_1, \ln f_2, \cdots \ln f_n|M_n,I) = \frac{n!}{[\ln (f_H/f_L)]^n}$  & 1/1.01 d & 1/1000 yr  \\
&\ ($n=$number of planets)  & &  \\
& & & \\
Velocity $K_i$  &  Modified Jeffreys~\footnote{Since the prior lower limits for $K$ and $s$ include zero, we used a modified Jeffreys prior of the form
\begin{equation}
p(X|M,I) = \frac{1}{X+X_0}\; \frac{1}{\ln\left(1+\frac{X_{\rm max}}{X_0}\right)}
\label{eq:orbit13}
\end{equation}
For $X \ll X_0$, $p(X|M,I)$ behaves like a uniform prior and for $X \gg X_0$ it behaves like a Jeffreys prior. The $\ln\left(1+\frac{ X_{\rm max}}{X_0}\right)$ term in the denominator ensures that the prior is normalized in the interval 0 to $X_{\rm max}$.} & 0 \ (K$_0 = 1)$ &  $K_{\rm max}\ \left(\frac{P_{\rm min}}{P_i}\right)^{1/3} \frac{1}{\sqrt{1-e_i^2}}$ \\
\ \ \  (m s$^{-1}$) & & & \\
  & \ \ \ \ \ $\frac{(K+K_0)^{-1}}{\ln{\left[1+\frac{K_{\rm max}}{K_0} \ \left(\frac{P_{\rm min}}{P_i}\right)^{1/3} \frac{1}{\sqrt{1-e_i^2}}\right]}}$
 &  & $K_{\rm max}=2129$\\
 & & & \\
V  (m s$^{-1}$) & $-K_{\rm max}$ & $K_{\rm max}$ &   \\
& & & \\
$e_i$ Eccentricity & Uniform & 0 & 1 \\
& & & \\
& & & \\
$\omega_i$ Longitude of & Uniform & $0$ & $2 \pi$ \\
\ \ \ \ periastron &  &  & \\
& & & \\
$s$ Extra noise   (m s$^{-1}$) & $\frac{(s+s_0)^{-1}}{\ln{\left(1+\frac{s_{\rm max}}{s_{0}}\right)}}$ & 0  \ (s$_0 = 1$)& $K_{\rm max}$  \\
\ \ \ (m s$^{-1}$) & & & \\
\hline
\end{tabular}
\end{minipage}
\end{table*}

\citealt{Gregory2007} discussed two different strategies to search the orbital frequency parameter space for a multi-planet model: (a) an upper bound on $f_1 \le f_2 \le \cdots \le f_n$  is utilized to maintain the identity of the frequencies, and (b) all $f_i$ are allowed to roam over the entire frequency range  and the parameters re-labeled afterwards. Case (b) was found to be significantly more successful at converging on the highest posterior probability peak in fewer iterations during repeated blind frequency searches. In addition, case (b) more easily permits the identification of two planets in 1:1 resonant orbits. We also adopted approach (b) in the current analysis. 

All of the models considered in this paper incorporate an extra noise parameter, $s$, that can allow for any additional noise beyond the known measurement uncertainties~\negthinspace\footnote{In the absence of detailed knowledge of the sampling distribution for the extra noise, we pick a Gaussian because, for any given finite noise variance, it is the distribution with the largest uncertainty as measured by the entropy, i.e., the maximum entropy distribution (\citealt{Jaynes1957}, \citealt{Gregorybook} section 8.7.4.)}. We assume the noise variance is finite and adopt a Gaussian distribution with a variance $s^2$. Thus, the combination of the known errors and extra noise has a Gaussian distribution with variance $= \sigma_i^2 + s^2$, where $\sigma_i$ is the standard deviation of the known noise for i$^{\mbox{\tiny th}}$ data point. For example, suppose that the star actually has two planets, and the model assumes only one is present. In regard to the single planet model, the velocity variations induced by the unknown second planet acts like an additional unknown noise term. Other factors like star spots and chromospheric activity can also contribute to this extra velocity noise term which is often referred to as stellar jitter. Several researchers have attempted to estimate stellar jitter for individual stars based on statistical correlations with observables (e.g., \citealt{Saar1997}, \citealt{Saar1998}, \citealt{Wright2005}). In general, nature is more complicated than our model and known noise terms. Marginalizing $s$ has the desirable effect of treating anything in the data that can't be explained by the model and known measurement errors as noise, leading to conservative estimates of orbital parameters (see Sections 9.2.3 and 9.2.4 of \citet{Gregorybook} for a tutorial demonstration of this point). If there is no extra noise then the posterior probability distribution for $s$ will peak at $s = 0$. The upper limit on $s$ was set equal to $K_{\rm max}$. We employed a modified Jeffrey's prior for $s$ with a knee, $s_0 = 1$m s$^{-1}$. 

\begin{figure}
\includegraphics[width=90mm]{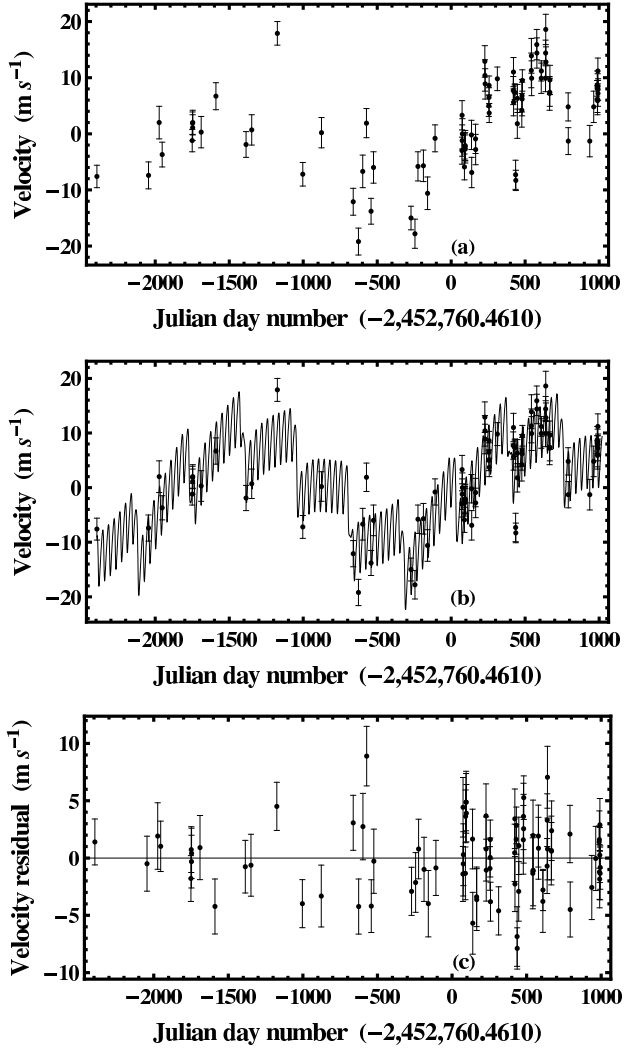}
 \caption{The data is shown in panel (a) and the best fitting three planet ($P_1 = 38$, $P_2 = 357$ , \& $P_3 = 1928$ days) model versus time is shown in (b). Panel (c) shows the residuals.}
\label{fig:dattimres}
\end{figure}

\section{Results}

Figure~\ref{fig:dattimres} shows the precision radial velocity data for HD 11964 from \citet{Butler2006} who reported a single planet with $M \sin i = 0.61\pm0.10$ in a $2110\pm270$ day orbit with an eccentricity of $0.06\pm0.17$. Panels (b) and (c) show our best fitting three planet light curve and residuals.

If we assume that all the models considered are equally probable {\it a priori}, then as shown in Section~\ref{sec:modsel}, the three planet model is $\ge 600$ times more probable than the next most probable model which is a two planet model. In this section, we mainly focus on the MCMC results for the three planet model.

\begin{figure*}
\includegraphics[width=180mm]{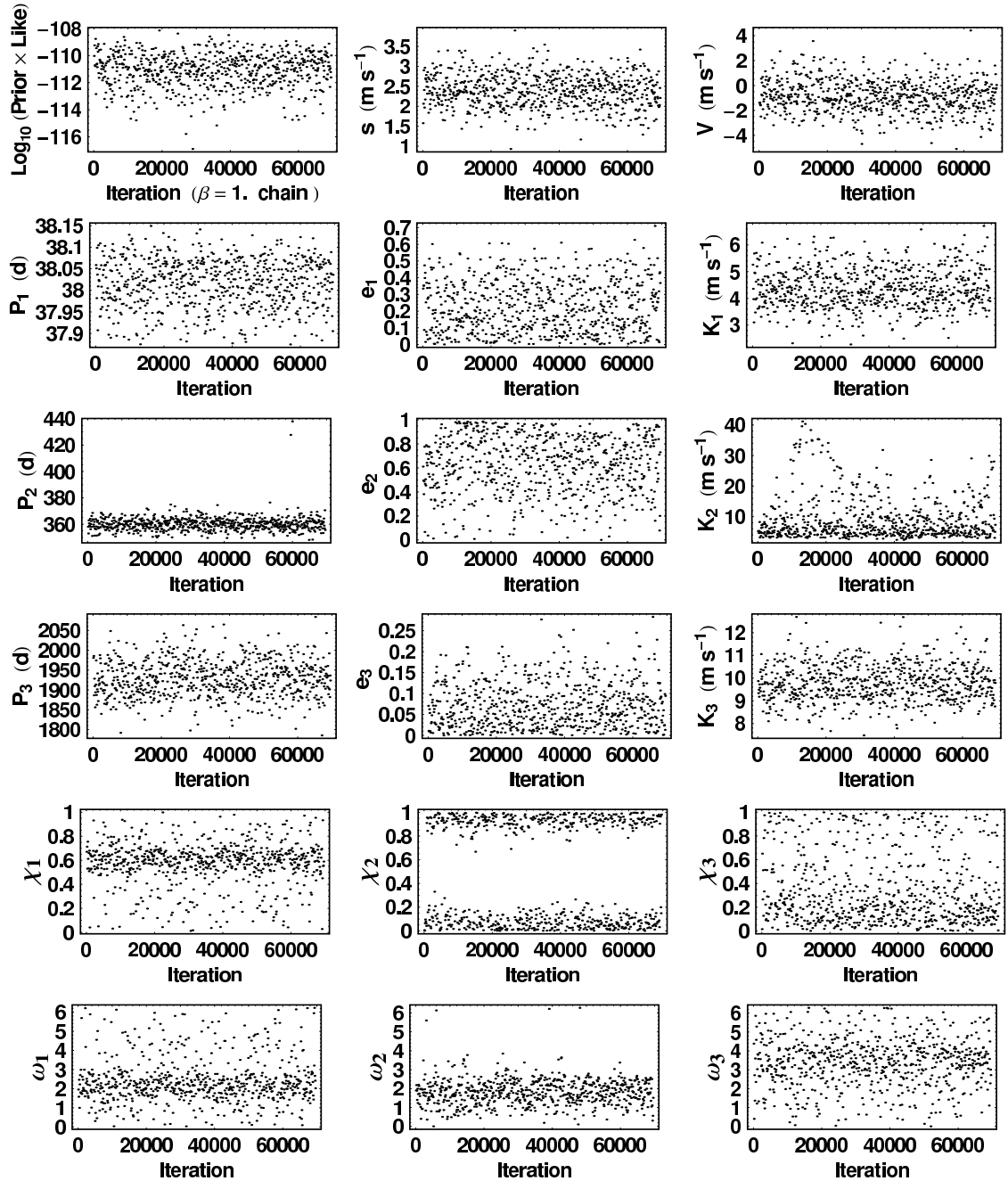}
 \caption{MCMC post burn-in parameter iterations for a three planet model. The upper left panel is a plot of Log$_{10}$(prior $\times$ likelihood).}
\label{fig:iter}
\end{figure*}

Figure~\ref{fig:iter} shows post burn-in MCMC iterations for the parameters of a three planet model returned by the Kepler periodogram, starting from an initial location ($P_1=10, P_2=500\ \& \ P_3=2300$d) in period parameter space. Similar results were obtained with other different starting positions. A total $10^6$ iterations were used with every tenth iteration stored. For display purposes only every hundredth stored point is plotted in the figure. The upper left panel is a plot of Log$_{10}$(prior $\times$ likelihood). The next two panels of the top row shows the extra noise parameter $s$ and the constant velocity parameter $V$, respectively. The remaining panels show the orbital parameters for each of the three periods. The equilibrium solution corresponds to $P_1=38$d, $P_2=360$d, \& $P_3=1924$d (for comparison, the two planet model MCMC yielded two solutions of $P_1=360\ \& \ P_2=1990$d and $P_2=38\ \& \ P_3=1932$d, respectively). All the traces appear to have achieved an equilibrium distribution. 

\begin{figure}
\includegraphics[width=90mm]{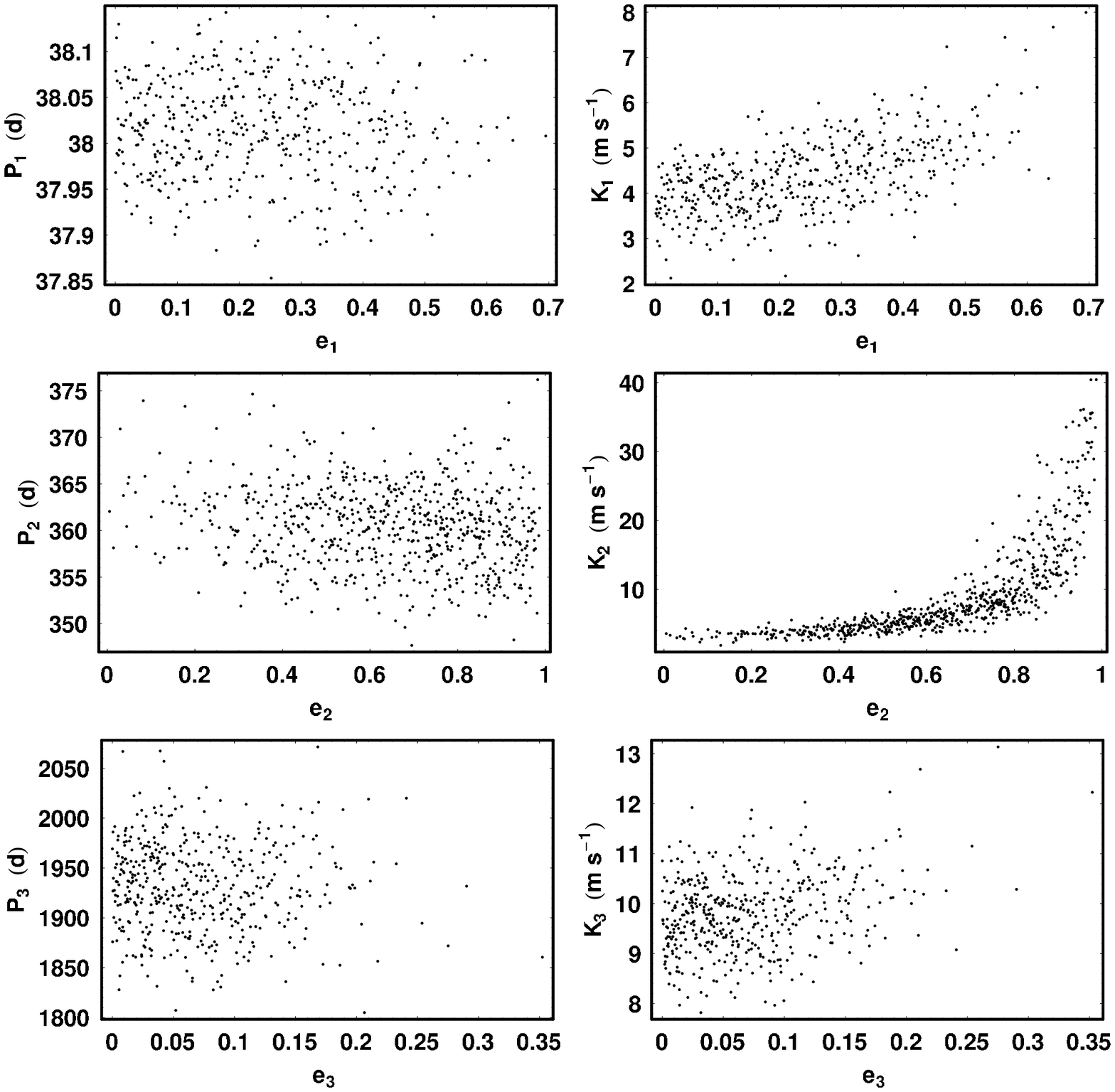}
\caption{A selection of joint marginal parameter probability distributions for the three planet model.}
\label{fig:corr}
\end{figure}

The $\chi_i$ and $\omega_i$ traces were derived from the corresponding $\psi_i,\phi_i$ traces. The $\psi_i,\phi_i$ traces are not shown. A correlation is clearly evident between $P_2$ and $e_2$ which is best seen in the joint marginals plotted in Figure~\ref{fig:corr}. Each dot is the result from one iteration.  

The \citet{Gel} statistic is typically used to test for convergence of the parameter distributions. In parallel tempering MCMC, new widely separated parameter values are passed up the line to the $\beta = 1$ simulation and are occasionally accepted. Roughly every 100 iterations the $\beta = 1$ simulation accepts a swap proposal from its neighboring simulation. 
The final $\beta = 1$ simulation is thus an average of a very large number of independent $\beta = 1$ simulations. What we have done is divide the $\beta = 1$ iterations into ten equal time intervals and inter-compared the ten different essentially independent average distributions for each parameter using a Gelman-Rubin test. For all of the three planet model parameters the Gelman-Rubin statistic was $\le 1.03$.

\begin{figure*}
\includegraphics[width=150mm]{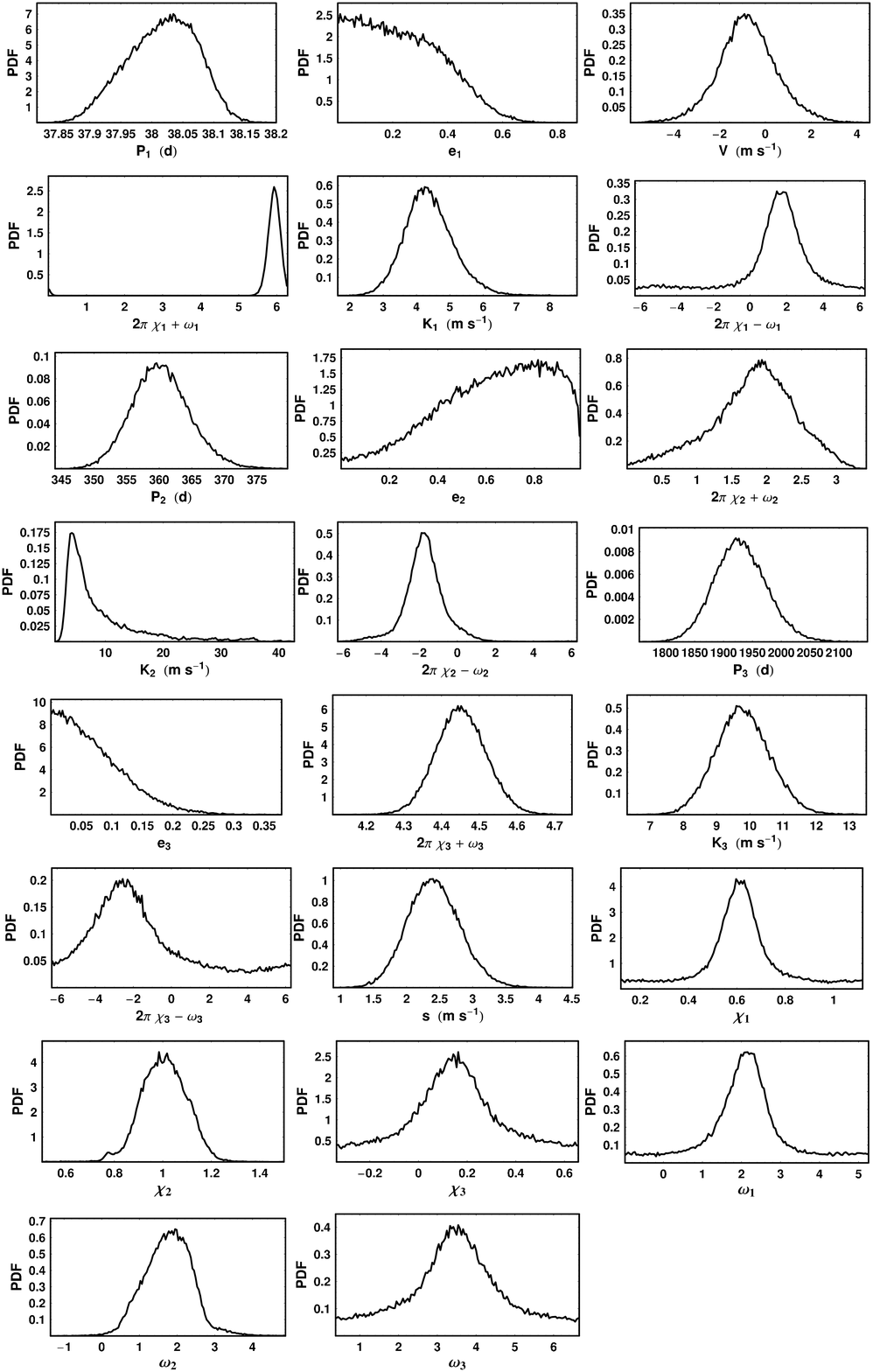}
\caption{Marginal parameter probability distributions for the three planet model.}
\label{fig:marg}
\end{figure*}

Figure~\ref{fig:marg} shows the individual parameter marginal distributions for the three planet model. Table~\ref{tab:parerrorsM2} gives our Bayesian three planet orbital parameter values and their errors. The parameter value listed is the median of the marginal probability distribution for the parameter in question and the error bars identify the boundaries of the 68.3\% credible region. The value immediately below in parenthesis is the maximum {\it a posteriori} (MAP) value determined using the \citet{NelderMead1965} downhill simplex method. The values derived for the semi-major axis and $M \sin i$, and their errors, are based on the assumed mass of the star $= 1.49\pm0.15$ M$_{\sun}$ \citep{Valenti2005}. \citet{Butler2006} assumed a mass of $= 1.12$ M$_{\sun}$ but also quote \citet{Valenti2005} as the reference. The last row gives the Bayesian estimate of the extra noise parameter (stellar jitter) for each model.

\begin{figure}
\includegraphics[width=90mm]{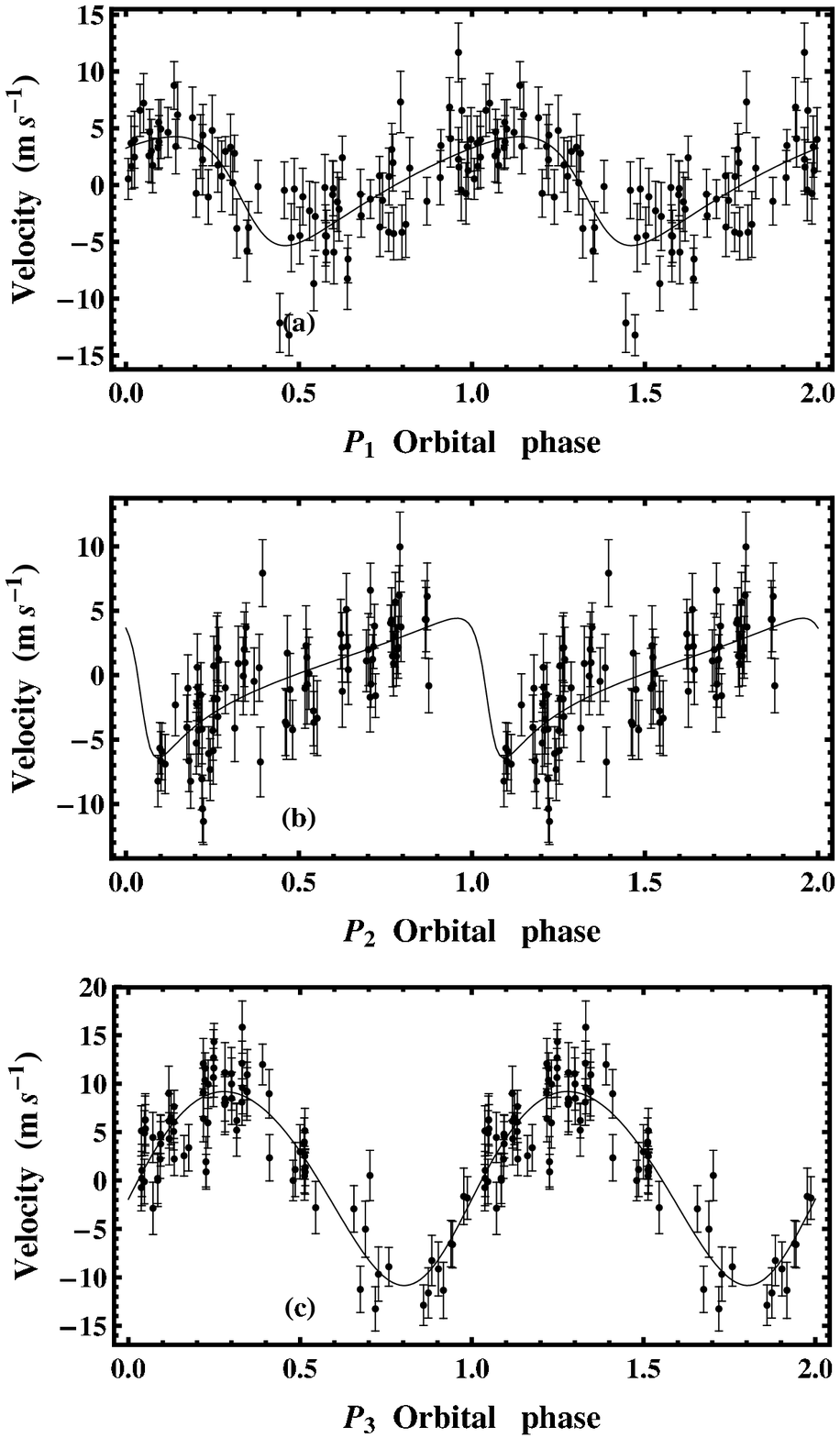}
 \caption{Panel (a) shows the data, with the best fitting $P_2$ and $P_3$ orbits subtracted, for two cycles of $P_1$ phase with the best fitting $P_1$ orbit overlaid. Panel (b) shows the data plotted versus $P_2$ phase with the best fitting $P_1$ and $P_3$ orbits removed. Panel (c) shows the data plotted versus $P_3$ phase with the best fitting $P_1$ and $P_2$ orbits removed. }
\label{fig:phase}
\end{figure}

In Figure~\ref{fig:phase}, panel (a) shows the data, with the best fitting $P_2$ and $P_3$ orbits subtracted, for two cycles of $P_1$ phase with the best fitting $P_1$ orbit overlaid. Panel (b) shows the data plotted versus $P_2$ phase with the best fitting $P_1$ and $P_3$ orbits removed. Panel (c) shows the data plotted versus $P_3$ phase with the best fitting $P_1$ and $P_2$ orbits removed.
\begin{table}
 \centering
 \begin{minipage}{140mm}
  \caption{Three planet model parameter estimates.}
  \label{tab:parerrorsM2}
  \begin{tabular}{@{}lllll@{}}
  \hline
   Parameter  & planet 1 & planet 2 & planet 3 \\
\hline
$P$  (d) & $38.02_{-0.5}^{+0.6}$ & $360_{-4}^{+4}$& $1925_{-44}^{+44}$  \\
& (38.07)& (357)& (1928)\\
& & &  \\
$K$ (m s$^{-1}$) & $4.3_{-0.7}^{+0.7}$ & $6.1_{-3.3}^{+3.0}$ & $9.7_{-0.8}^{+0.8}$  \\
& (4.8) & (5.4) & (10.0) \\
& & &  \\
$e$ & $0.23_{-.22}^{+.10}$ & $0.63_{-.13}^{+.35}$  & $0.05_{-.05}^{+.03}$   \\
& (0.31) & (0.63) & (0.09) \\
& & &  \\
$\omega$  (deg) & $123_{-48}^{+41}$ & $103_{-34}^{+38}$ &  $195_{-74}^{+80}$  \\
& (111) & (107) & (205) \\
& & &  \\
$a$  (au) & $0.2527_{-.0085}^{+.0085}$ & $1.132_{-.039}^{+.039}$ &  $3.46_{-.13}^{+.13}$  \\
& (0.253) & (1.124) &  (3.46)  \\

& & & & \\
$M \sin i$  ($M_J$) & $0.090_{-.015}^{+.014}$ & $0.213_{-.067}^{+.058}$ &  $0.77_{-.08}^{+.08}$ \\
& (0.098) & (0.191) &  (0.795) \\
& & &  \\
Periastron & $12737_{-3}^{+6}$ & $12397_{-32}^{+35}$ &  $10535_{-414}^{+401}$  \\
\ passage &  (12736) & (12421)&  (10564) \\
\ (JD - 2,440,000) & & &  \\
& & &  \\
$s$ (m s$^{-1}$) & $4.9_{-.5}^{+.5}$ & $3.7_{-.4}^{+.4}$  & $2.4_{-.4}^{+.4}$   \\
& (4.7) & (3.3) & (1.9)  \\
\hline
\end{tabular}
\end{minipage}
\end{table}

\section{Model selection}
\label{sec:modsel}

To compare the posterior probabilities of the i$^{\rm th}$ planet model to the one planet models we need to evaluate the odds ratio,
$O_{i1} =p(M_{i} | D,I)/p(M_{1} | D,I)$, the ratio of the posterior probability
of model $M_{i}$ to model $M_{1}$.  Application of Bayes's 
theorem leads to,
\begin{equation}
O_{i1} = {p(M_{i} | I) \over p(M_{1} | I)}\;
      {p(D | M_{i},I) \over p(D | M_{1},I)}
       \equiv {p(M_{i} | I) \over p(M_{1} | I)}\; B_{i1}
\label{eq:orbit22}
\end{equation}
where the first factor is the prior odds ratio, and the second factor
is called the {\it Bayes factor}. The Bayes factor is the ratio of
the marginal (global) likelihoods of the models. The MCMC algorithm produces
samples which are in proportion to the posterior probability distribution which is fine for parameter
estimation but one needs the proportionality constant for estimating the model marginal likelihood. 
\citet{Clyde2006} recently reviewed the state of techniques for model selection from a statistics perspective and \citet{FordGregory2006} have evaluated the performance of a variety of marginal likelihood estimators in the extrasolar planet context. 

In this work we will compare the results from three marginal likelihood estimators: (a) parallel tempering, (b) ratio estimator, and (c) restricted Monte Carlo. 
A brief outline of each method is presented in Sections~\ref{sec:partempML}, \ref{sec:reML}, and \ref{sec:RMC}. The results are summarized in Section~\ref{sec:summML}.

\subsection{Parallel tempering estimator}
\label{sec:partempML}

\begin{figure}
\includegraphics[width=80mm]{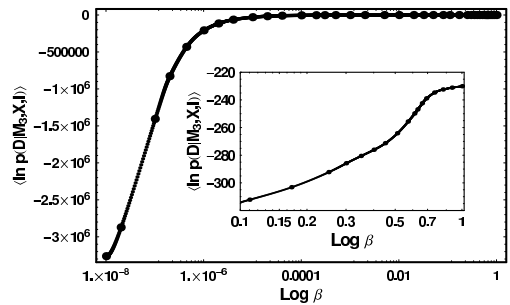}
 \caption{A plot of $\langle \ln[p(D|M_{3},X,I)]\rangle_{\beta}$ versus $\beta$ for the three planet model. The inset shows a blow-up of the range $\beta= 0.1$ to 1.0.}
\label{fig:bayesfactor}
\end{figure}

The MCMC samples from all ($n_\beta$) simulations can be used to calculate the marginal likelihood of a model
according to equation~(\ref{eq:Zbeta10}) \cite{Gregorybook}.
\begin{equation}
\ln[p(D|M_{i},I)] =  \int d\beta \langle \ln[p(D|M_{i},\vec{X},I)]\rangle_{\beta},
\label{eq:Zbeta10}
\end{equation}
where $i = 0,1,2,3,4$ corresponds to the number of planets, and $\vec{X}$ represent a vector of the model parameters which includes the extra Gaussian noise parameter $s$. In words, for each of the $n_\beta$ parallel simulations, compute the expectation value (average) of the natural logarithm of the likelihood for post burn-in MCMC samples. 
It is necessary to use a sufficient number of tempering levels that we can estimate the above integral by interpolating values of 
\begin{equation}
\langle \ln[p(D|M_{i},\vec{X},I)]\rangle_{\beta}= \frac{1}{n}\sum_t \ln[p(D|M_{i},\vec{X},I)]_{\beta},
\label{eq:Zbeta10a}
\end{equation}
in the interval from $\beta = 0$ to 1, from the finite set. For this problem we used 40 tempering levels in the range $\beta = 10^{-8}$ to 1.0. Figure~\ref{fig:bayesfactor} shows a plot of $\langle \ln[p(D|M_{i},\vec{X},I)]\rangle_{\beta}$ versus $\beta$. The inset shows a blow-up of the range $\beta= 0.1$ to 1.0. 

The relative importance of different decades of $\beta$ can be judged from Table~\ref{tab:FracError}.
The second column gives the fractional error that would result if this decade of $\beta$ was not included and thus indicates the sensitivity of the result to that decade. The fractional error falls rapidly with each decade and for the lowest decade explored in this run, $\beta = 10^{-8}\ - \ 10^{-7}$, reaches $0.21$. From Figure~\ref{fig:bayesfactor}, it is apparent that the steep drop in the curve that occurs below $\beta = 10^{-6}$ shows a significant change in curvature in the direction of leveling off similar to that experienced in the case of HD 208784 \citep{Gregory2007} and HD 188133 \citep{FordGregory2006}. In the case of HD 208487 (2 planet model), the fractional error reached 0.16 in the range $\beta = 10^{-6} - 10^{-7}$ and the error fell to 0.02 in the next decade. For HD 188133 (one planet model), the fractional error reached 0.26 in the range $\beta = 10^{-5} - 10^{-6}$ and the contribution to the fractional error for the next 4 decades was 0.14. Based on these comparisons we estimate that ignoring lower decades of $\beta$ will result in a systematic underestimate of $p(D|M_3,I)$ of $\sim 15 \%$. A similar table for the two planet PT results gave the fractional error for the lowest decade at $0.11$. 
\begin{table}
 \centering
 \begin{minipage}{80mm}
  \caption{Fractional error versus $\beta$ for the three planet model results shown in Figure~\ref{fig:bayesfactor}.}
  \label{tab:FracError}
  \begin{tabular}{@{}lllll@{}}
  \hline
   $\beta$ range    &    Fractional error \\
\hline
$1.0\ -\ 10^{-1}$      & $9.59 \times 10^{101}$  \\
$10^{-1}\ -\ 10^{-2}$  & $4.55 \times 10^{13}$  \\
$10^{-2}\ -\ 10^{-3}$  & $225$  \\
$10^{-3}\ -\ 10^{-4}$  & $3.46$  \\
$10^{-4}\ -\ 10^{-5}$  & $1.32$  \\
$10^{-5}\ -\ 10^{-6}$  & $0.77$  \\
$10^{-6}\ -\ 10^{-7}$  & $0.51$  \\
$10^{-7}\ -\ 10^{-8}$  & $0.21$  \\
\hline
\end{tabular}
\end{minipage}
\end{table}

\subsection{Marginal likelihood ratio estimator}
\label{sec:reML}

Our second method \footnote{Initially proposed by J. Berger, at an Exoplanet Workshop sponsored by the Statistical and Applied Mathematical Sciences Institute in Jan. 2006} was introduced by \citet{FordGregory2006}. It makes use of an additional sampling distribution $h(\vec{X})$. Our starting point is Bayes' theorem
\begin{equation}
p(\vec{X}|M_{i},I) = \frac{p(\vec{X}|M_i,I) p(D|M_i,\vec{X},I)}{p(D|M_i,I)}.
\label{eq:marglike1}
\end{equation}
Re-arranging the terms and multiplying both sides by $h(\vec{X})$ we obtain
\begin{eqnarray}
& & p(D|M_i,I) p(\vec{X}|M_{i},I) h(\vec{X}) = \nonumber \\
& & \ \ \ \ \ \ \ \ \ \ \ p(\vec{X}|M_i,I) p(D|M_I,\vec{X},I) h(\vec{X}).
\label{eq:marglike2}
\end{eqnarray}
Integrate both sides over the prior range for $\vec{X}$.
\begin{eqnarray}
& & p(D|M_i,I)_{RE} \int p(\vec{X}|M_{i},I) h(\vec{X}) d\vec{X} =  \nonumber \\
& & \ \ \ \ \ \ \ \ \ \ \int p(\vec{X}|M_i,I) p(D|M_I,\vec{X},I) h(\vec{X}) d\vec{X}.
\label{eq:marglike3}
\end{eqnarray}
The ratio estimator of the marginal likelihood, which we designate by $p(D|M_i,I)_{RE}$, is given by 
\begin{equation}
p(D|M_{i},I)_{RE} = \frac{\int p(\vec{X}|M_i,I) p(D|M_i,\vec{X},I) h(\vec{X}) d\vec{X}}{\int p(\vec{X}|M_{i},I) h(\vec{X}) d\vec{X}}.
\label{eq:marglikeRE}
\end{equation}

To obtain the marginal likelihood ratio estimator, $p(D|M_{i},I)_{RE}$, we approximate the numerator by drawing samples $\tilde{X}^1,\tilde{X}^2,\cdots,\tilde{X}^{n_s^{\prime}}$ from $h(\vec{X})$ and approximate the denominator by drawing samples $\vec{X}^1,\vec{X}^2,\cdots,\vec{X}^{n_s}$ from the $\beta = 1$ MCMC post burn-in iterations.
\begin{equation}
p(D|M_{i},I)_{RE} = \frac{\frac{1}{n'_s} \sum_{i=1}^{n'_s} p(\tilde{X}^i|M_i,I) p(D|M_i,\tilde{X}^i,I)}
{\frac{1}{n_s} \sum_{i=1}^{n_s} h(\vec{X}^i)} \,.
\label{eq:marglikeRE1}
\end{equation}
The arbitrary function $h(\vec{X})$ was set equal to a multivariate normal distribution (multinormal) with a covariance matrix equal to twice the covariance matrix computed from a sample of the $\beta=1$ MCMC output. We used~\footnote{According to \citet{FordGregory2006}, the numerator converges more rapidly than the denominator.} $n_s^\prime = 10^5$ and $n_s$ from $10^4$ to $2 \times 10^5$. Some of the samples from a multinormal $h(\vec{X})$ can have non physical parameter values (e.g. $K <0$). Rejecting all non physical samples corresponds to sampling from a truncated multinormal. The factor required to normalize the truncated multinormal is just the ratio of the total number of samples from the full multinormal to the number of physical valid samples. Of course we need to use the same truncated multinormal in the denominator of equation~(\ref{eq:marglikeRE}) so the normalization factor cancels. $p(D|M_{2},I)_{RE}$ converges much more rapidly than the parallel tempering estimator \cite{Gregory2007} and the parallel tempering estimator, $p(D|M_{2},I)_{PT}$, required 40 $\beta$ simulations instead of one.
  
\subsubsection{Mixture model}

It is clear that a single multinormal distribution can not be expected to do a very good job of representing the correlation between the parameters that is evident between $P_2$ and $e_2$ in Figure~\ref{fig:corr}. Following \citet{FordGregory2006}, we improve over the single multinormal by using a mixture of multivariate normals by setting 
\begin{equation}
h({\vec{X}}) = \frac{1}{n_c} \sum_{j=1}^{n_c} h_j({\vec{X}})
\label{eq:mix}
\end{equation}
where we must determine a covariance matrix for each $h_j({\vec{X}})$ using the
posterior sample. We choose each mixture component to be a multivariate normal distribution, 
$
h_j(\vec{X}) = N(\vec{X} |\vec{X}_j, \Sigma_j )
$,
where we must determine a covariance matrix for each $h_j$ using the
posterior sample.   First, we compute $\vec{\rho}$, defined to be a
vector of the sample standard deviations for each of the components of
$\vec{X}$, using the posterior sample.  Next, define the
distance between the posterior sample $\vec{X}_i$ and the center
of $h_j(\vec{X})$,
$
d^2_{ij} = \sum_{k} \left(\vec{X}_{ki} - \vec{X}_{kj} \right)^2 /
\rho_k^2
$,
where $k$ indicates the element of $\vec{X}$ and $\vec{\rho}$.
Now draw another random subset of $100 n_c$ samples from the
original posterior sample (without replacement), select the
$100$ posterior samples closest to each mixture component and
use them to calculate the covariance matrix, $\Sigma_j$, for each
mixture component. Since the posterior sample is assumed to have
fully explored the posterior, $h(\vec{X})$ should be quite
similar to the posterior in all regions of significant probability,
provided that we use enough mixture components.

\subsection{Restricted Monte Carlo marginal likelihood estimate}
\label{sec:RMC}

We can also make use of Monte Carlo integration to evaluate the marginal likelihood as given by equation~(\ref{eq:marglike4}).
\begin{equation}
p(D|M_i,I) = \int p(X|M_i,I) p(D|M_I,X,I) dX.
\label{eq:marglike4}
\end{equation}
Monte Carlo (MC) integration can be very inefficient in exploring the whole prior parameter range, but once we have established the significant regions of parameter space with the MCMC results, this is no longer the case.
The outer borders of the MCMC marginal parameter distributions were used to delineate the boundaries of the volume of parameter space to be used in the Monte Carlo integration. RMC integration was carried out for models $M_1$, $M_2$, and $M_3$ based on $4 \times 10^6$ samples and repeated three times. 

\subsection{Summary of model selection results}
\label{sec:summML}

Table~\ref{tab:modelSel} summarizes the marginal likelihoods and Bayes factors comparing models 
$M_{0}$, $M_{2}$, $M_3$, $M_4$ to $M_{1}$. For model $M_{0}$, the marginal likelihood was obtained by numerical integration.  For $M_{1}$, the value and error estimate are based on the RMC method discussed in Section~\ref{sec:RMC}, the RE method (1 mixture component), discussed in Section~\ref{sec:reML}, and the RE method (100 mixture components). Each method was repeated 3 times on the same posterior sample to ascertain the variance of repeated trials. The quoted uncertainty is the standard deviation of the repeats. The sample error of the mean is a factor of $1/\sqrt{3}$ smaller. Since all three methods yield approximate marginal likelihoods it is not clear which is the most accurate but we are inclined to favor the RE method with 100 mixture components. All three estimates agree within 15\%. 

For model $M_2$, two peaks in the joint posterior probability distribution were detected: (A) $P_1 = 362$d, $P_2 =1984 $d, and (B) $P_1 = 37.98$d, $P_2 = 1897 $d. The contribution to the marginal likelihood from each peak was estimated from the posterior samples from the $\beta = 1$ chain after filtering the posterior samples in $P_1$ and $P_2$ to exclude samples from the other peak. The ratio estimator method was employed with three different mixture components 1, 100, \& 500. For each peak, the results agreed well within a factor of better than 2. The 100 and 500 mixture components agreed more closely and appeared to be systematically lower than for the 1 component version. On the basis of these results, peak A is a factor of $\sim 10$ more probable. Combining the RE 500 component results for both peaks yields a $p(D|M_3,I) = 2.3 \times 10^{-124}$ which is close to the $3.0 \times 10^{-124}$ value derived from the parallel tempering method which was discussed in Section~\ref{sec:partempML}. Our final estimate is $(2.5\pm 0.5)\times 10^{-124}$.

\begin{table*}
 \centering
  \caption{Marginal likelihood estimates, Bayes factors and probabilities for the 5 models. The last two columns list the MAP value of extra noise parameter, $s$, and the RMS residual.}
  \label{tab:modelSel}
  \begin{tabular}{@{}llllllcc@{}}
  \hline
   Model & Method & Mixture& Marginal & Bayes factor & Probability & $s$ &RMS residual \\
         & & components & Likelihood & \ \ nominal & \ \ nominal & (m s$^{-1})$ & (m s$^{-1}$)\\
\hline
$M_{0}$  & Exact & & $6.86 \times 10^{-138}$ & $2.7 \times 10^{-10}$ & $2.5 \times 10^{-18}$ & 7.6 &8.0\\
& & & & & & &\\
$M_{1}$ & RMC & & $(2.31\pm 0.01)\times 10^{-128}$ &  &  & &\\
$M_{1}$  & RE & 1 & $(2.86\pm 0.07)\times 10^{-128}$ &  & & &\\
$M_{1}$  & RE & 100 & $(2.50\pm 0.06)\times 10^{-128}$ &  & & &\\
$M_{1}$  & Summary &  & $(2.50\pm 0.4)\times 10^{-128}$ & 1.0 & $9 \times 10^{-9}$ & 4.7 & 5.3\\
& & & & & &\\
$M_{2A}$  & RMC & & $(1.5\pm 0.2)\times 10^{-124}$ &  &  & &\\
$M_{2A}$  & RE  & 1 & $ (3.02\pm 0.16) \times 10^{-124}$ &  & & &\\
$M_{2A}$  & RE  & 100 & $ (2.08\pm 0.06)\times 10^{-124}$ & & & &\\
$M_{2A}$  & RE  & 500 & $(1.98\pm 0.08)\times 10^{-124}$ &  & & &\\
$M_{2B}$  & RMC & & $(3.3\pm 0.2)\times 10^{-125}$ &  &  & &\\
$M_{2B}$  & RE  & 1 & $ (3.86\pm 0.26) \times 10^{-125}$ &  & & &\\
$M_{2B}$  & RE  & 100 & $ (3.28\pm 0.15) \times 10^{-125}$ & & & &\\
$M_{2B}$  & RE  & 500 & $(2.95\pm 0.18)\times 10^{-125}$ &  & & &\\
$M_{2}$  & RE (A+B) & 500 & $(2.3\pm 0.08)\times 10^{-124}$ &  & & &\\
$M_{2}$  & RMC (A+B) & & $(1.8\pm 0.3)\times 10^{-124}$ &  & & &\\
$M_{2}$  & PT &  &  $(3^{\times 2}_{\times 1/2})  10^{-124}$ &  & & &\\
$M_{2}$  & Summary &  &  $(2.5\pm 0.5)\times 10^{-124}$ & $1.0 \times 10^{4}$ & $9 \times 10^{-5}$ & 3.3 & 4.1\\
& & & & & & &\\
$M_{3}$  & RMC &  & $(1.8\pm 0.3) \times 10^{-121}$ &  & & &\\
$M_{3}$  & RE & 1 & $(14\pm 3) \times 10^{-119}$ &  & & &\\
$M_{3}$  & RE & 100 & $(4.95\pm 0.44)\times 10^{-119}$ &  & & &\\
$M_{3}$  & RE & 500 & $(5.00\pm 0.44)\times 10^{-119}$ &  & & &\\
$M_{3}$  & PT &  & $ 2.8\times 10^{-120}$ &  & & &\\
$M_{3}$  & Summary &  & $(2.8^{\times 18}_{\times 1/16})\times 10^{-120}$ &  $1.1 \times 10^{8}$ & 0.99991 & 1.9 & 3.0\\
& & & & & & &\\
$M_{4}$  & RE & 100 & $ 3.2 \times 10^{-126}$ &  & & &\\
$M_{4}$  & RE & 100 & $ 1.5 \times 10^{-125}$ &  & & &\\
$M_{4}$  & RE & 100 & $ 3.7 \times 10^{-125}$ &  & & &\\
$M_{4}$  & RE & 100 & $ 1.0 \times 10^{-124}$ &  & & &\\
$M_{4}$  & Summary &  & $\le 1.9 \times 10^{-125}$ &  $\le760$ & $M_4$ excluded & 1.5 & 2.5\\
\hline
\end{tabular}
\end{table*}

The marginal likelihood for $M_3$ was estimated from the posterior samples from the $\beta = 1$ chain using the ratio estimator (RE), for 1, 100, and 500 mixture components, by RMC integration, and by the PT method which makes use of the samples from 40 tempering chains. The results for the 100 and 500 mixture components are in good agreement and are a factor of $\sim 2$ less than the one component RE results. It is to be expected that the multiple mixture component versions will do a better job of modeling correlated parameters than a single component model. 

\begin{figure*}
\includegraphics[width=140mm]{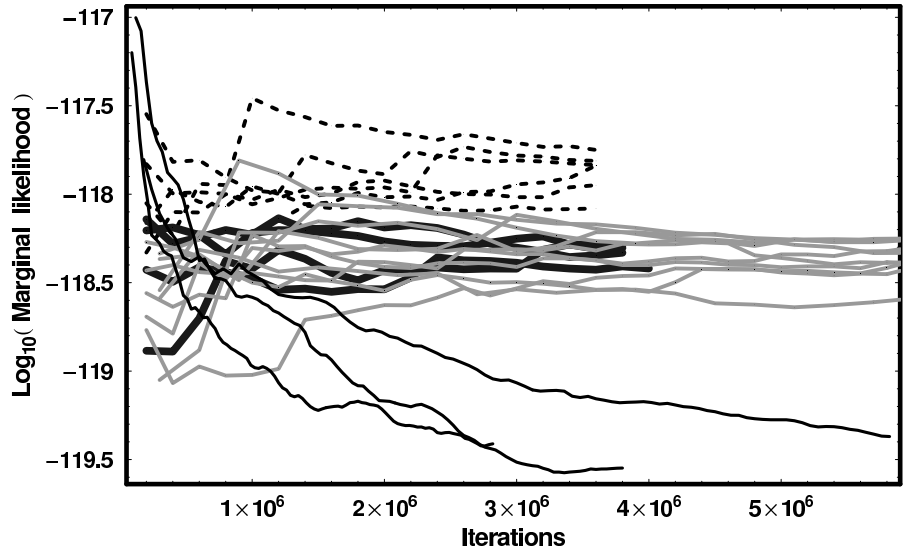}
 \caption{The thin solid black curves show 3 repeats of the parallel tempering marginal likelihood method versus iteration number for the three planet model. The dashed curves show 6 repeats of the ratio estimator method using only one mixture component. The thick black curves shows the result for 4 repeats of the ratio estimator method using 100 mixture components and the gray curves correspond to trials using 500 mixture components.}
\label{fig:bayesfactor1}
\end{figure*}

Figure~\ref{fig:bayesfactor1} shows the behavior of the PT marginal likelihood estimator when it is computed using different numbers of iterations taken from a particular run. The results for the three planet model, using three such MCMC runs, are shown by the thin black curves. The iteration number indicated on the abscissa is the total number of iterations executed but only a fraction (typically every tenth or less) were saved and used for this analysis. For comparison, the results from repeated trials of the RE marginal likelihood estimates versus iteration are shown. The dashed curves are RE using one component. The thick black curves are RE with 100 mixture components and the gray curves correspond to 500 components. It is apparent that two of the PT runs have not yet converged. The third appears to have leveled off at a value of $p(D|M_3,I) = 2.8 \times 10^{-120}$ but of course more iterations would be desirable. All of the PT results argue for a value significantly less than for the RE method. Finally, the result of two RMC trials is even lower at $1.8 \times 10^{-121}$. It is a difficult question to decide which estimate is best. As a summary we have quoted the PT result with errors that span the other two methods. In spite of the large uncertainty, the evidence favoring the three planet model is very strong. Assuming equal model priors, then from the lower limit for $p(D|M_3,I)$ of $1.8 \times 10^{-121}$ and the upper limit for $p(D|M_2,I)$ of $3.0 \times 10^{-124}$, we conclude that for a fair bet the odds in favor of $M_3$ over $M_2$ is $\ge 600$. Similarly, the odds in favor of $M_3$ over $M_1$ is $\ge 6 \times 10^{6}$. 

Table~\ref{tab:modelSel} also gives an estimate for $p(D|M_4,I)$ based on four repeats of the ratio estimator with 100 components. Because of the large spread in results, our summary is the geometric mean of the individual values. In view of the results for the three planet model we consider the RE estimates for $p(D|M_4,I)$ as an upper limit. 

Column 5 of Table~\ref{tab:modelSel} gives the nominal Bayes factor comparing each model to the one planet model. Assuming equal model priors, the probability of model $M_i$ is given by
\begin{equation}
p(M_{i}|D,I)=\frac{p(D|M_{i},I)}{\sum_{j=0}^3 p(D|M_{j},I)}. 
\label{eq:pd}
\end{equation}
Nominal model probabilities excluding model $M_4$ are given in Column 6. 
The results overwhelmingly favor the three planet model. 

Column 7 and 8 list the most probable values of the extra noise parameter, $s$, and the RMS residuals in m s$^{-1}$, respectively.

\section{Discussion}

In this paper, we have demonstrated that a sophisticated Bayesian analysis of the published data for HD 11964 finds strong evidence for two additional planets. Is it likely that there are many other cases among the ~200 published RV data sets that this type of an analysis would yield evidence for additional planets, or is the HD 11964 system likely to be unique or rare? To date,
the Bayesian MCMC Kepler periodogram has been run on only a small number of data sets including HD 73526 \citet{Gregory2005b} and HD 208487 \citet{Gregory2007}, which both yielded evidence for an additional planet. It thus appears likely that the algorithm is capable of detecting many additional exoplanet cantidates in the published RV data. Although the current implementation of the algorithm is not particular fast (19h for a typical 3 planet model run of $10^6$ iterations with 12 tempering chains), it has many advantages that were outlined in Section~\ref{sec:introduction}. 

One source of error in the measured velocities is "jitter", which is due in part to flows and inhomogeneities on the stellar surface. Wright (2005) gives a model that estimates, to within a factor of roughly 2 \citep{Butler2006}, the jitter for a star based upon a star's activity, color, Teff, and height above the main sequence. For HD 11964, \citet{Butler2006} quote a jitter estimate of $5.7$m s$^{-1}$, based on Wright's model. Our models $M_0$ to $M_4$ employ instead an extra Gaussian noise nuisance parameter, $s$, with a prior upper bound of equal to $K_{\rm max}=2129$m s$^{-1}$. Anything that cannot be explained by the model and published measurement uncertainties (which do not include jitter) contributes to the extra noise term. Of course, if we are interested in what the data have to say about the size of the extra noise term then we can readily compute the marginal posterior for $s$. The marginal for $s$ for $M_3$ is shown in the middle panel of the 6$^{\rm th}$ row in Figure~\ref{fig:marg}. The marginal for $s$ shows a pronounced peak with a median of $2.4$m s$^{-1}$ and a MAP value of $2.0$m s$^{-1}$. The MAP value of $s$ for all our models is tabulated in Table~\ref{tab:modelSel}. For $M_1$, the Map value is $4.9$m s$^{-1}$ which is well within the factor of two uncertainty of the jitter estimate given in \citet{Butler2006} based on Wright's model. The results of our Bayesian model selection analysis indicate that a three planet model is $\ge 6 \times 10^{6}$ times more probable than a one planet model with the previously estimated jitter.

It is interesting to compare the performance of the three marginal likelihood estimators employed in this work to their performance in the two planet fit for HD 208487 \citep{Gregory2007}. For HD 208487 the parallel tempering estimator, based on 34 chains, required $\sim 1.5 \times 10^6$ iterations for convergence. The two separate runs agreed within a factor of 2.2. The average of the two HD 208487 PT results agreed with the RMC and RE (one component) within $20\%$. For the two planet fit of HD 11964 with 40 chains, convergence required $5 \times 10^{6}$ iterations. Since there were two peaks in the posterior, the RE and RMC had to be run on each peak separately and the two peak contributions added before comparing to the PT result which integrates over the entire posterior. The single PT run agreed with the RMC and RE estimates within a factor of $\sim 2$. For both HD208487 and HD 11964, the RMC and RE results for a two planet model agreed within $\sim 25\%$.

For HD 11964, the model $M_3$ results from the three methods spanned a much larger range. Further it is clear that two of the three PT runs have not converged, in one case after $6 \times 10^6$ iterations with 40 tempering chains which took 16 days on a fast single core PC. This experience suggests that it may not be feasible to compute parallel tempering marginal likelihoods for models involving three or more planets, which will typically require $\ge 40$ chains. Parallel computing could help but there is still a need for more efficient, accurate and well calibrated methods for computing the marginal likelihoods. MCMC is great for parameter estimation, so perhaps more effort is required to include the number of planets as an additional parameter as has been done in other areas.    

\section{Conclusions}

In this paper, we further demonstrated the capabilities of an automated Bayesian parallel tempering MCMC approach to the analysis of precision radial velocities. The method is called a Bayesian Kepler periodogram because it is ideally suited for detecting signals that are consistent with Kepler's laws. However, it is more than a periodogram because it also provides full marginal posterior distributions for all the orbital parameters that can be extracted from radial velocity data. Moreover, it is a very general algorithm that can be applied to many other nonlinear model fitting problems.

The HD 11964 data \citep{Butler2006} has been re-analyzed using 1, 2, 3 and 4 planet models. The most probable model exhibits three periods of $38.02_{-0.05}^{+0.06}$, $360_{-4}^{+4}$ and $1924_{-43}^{+44}$ d, and eccentricities of $0.22_{-0.22}^{+0.11}$, $0.63_{-0.17}^{+0.34}$ and $0.05_{-0.05}^{+0.03}$, respectively. Assuming the three signals (each one consistent with a Keplerian orbit) are caused by planets, the corresponding limits on planetary mass ($M \sin i$) and semi-major axis are \\ 
$(0.090_{-0.14}^{+0.15} M_J$, $0.253_{-0.009}^{+0.009}\rm{au})$, $(0.21_{-0.07}^{+0.06} M_J$, $1.13_{-0.04}^{+0.04}\rm{au})$, $(0.77_{-0.08}^{+0.08} M_J$, $3.46_{-0.13}^{+0.13}\rm{au})$,\\ respectively. Based on our three planet model results, the remaining unaccounted for stellar jitter is $\sim 2$m s$^{-1}$. The small difference ($1.3 \sigma$) between the 360 day period and one year raise some concern about a possible instrumental effect and we suggest that it might be worth investigating the barycentric correction for the HD 11964 data.
 
Considerable attention was paid to the topic of Bayesian model selection. For model fitting involving $\le 2$ planets, all three marginal likelihood estimators were in good agreement. For a three planet fit, the RMC and RE results differed by a factor of $\sim 300$ and each differed from the PT result by a factor of $\sim 17$. Further improvements on the model selection side of this problem are clearly needed, requiring the development of more efficient, accurate and well calibrated methods for computing the marginal likelihoods.

\section*{Acknowledgments}

The author would like to thank John Skilling for suggesting an improvement to the sampling during the reparameterization stage of the algorithm. This research was supported in part by grants from the Canadian Natural Sciences and Engineering Research Council at the University of British Columbia.

\bsp

\label{lastpage}

\end{document}